\documentclass{article}
\title{Horizons in 2+1 dimensional collapse of particles}
\author{Dieter Brill\thanks{Department of Physics, University of Maryland College Park, MD 20782, USA},
Puneet Khetarpal\thanks{Rensselaer Polytechnic Institute, Troy, NY
12180, USA}, and Vijay Kaul$^*$}
\begin{document}
\maketitle

\section{Introduction}

Event horizons are generated by null geodesics. They are therefore
natural candidates for applying a time development equation of the
Raychaudhuri type. But in the present context we focus on the
non-smooth regions of horizons and other features that we describe
by geometrical construction, a kind of virtual use of the Raychaudhuri
equation. We hope that our material is nevertheless of interest to
readers of this volume.

The presence of an event horizon is the defining characteristic of
a black hole. When a black hole is formed by
gravitational collapse an event horizon starts at some stage of
the time development, spreads out increasing its area until it
encompasses all the dynamical features, and eventually becomes
stationary. The final black hole is then one of the small family
of ``hairless" types. The most active and interesting period in the
life of a black hole is the time near the formation of the
horizon. Because the horizon is a global property of the spacetime
and cannot be characterized locally, this interesting period is
typically studied only in numerically generated spacetime regions
that extend over a long time.

In two space- and one time-dimensions (2+1 D) the situation is
much simplified for several reasons. There is a simple kind of
matter that can collapse gravitationally and form black holes,
namely point particles; the geometry of spacetimes with such
collapsing matter is known exactly, at least in principle; and
there are no gravitational waves emitted in the collapse that may
allow the final black hole state to be reached only asymptotically
in time. Thus, in the collapse of 2+1~D particles, the formative
stage of the horizon lasts only a finite time and is over as soon
as the horizon has passed all the particles. These simplifying
features in a 2+1~D spacetime of course make it somewhat unrealistic as a
model for 3+1 D collapse, but one would still expect that many of
the lessons one learns in 2+1 dimensions survive, in some form, in
3+1 dimensions.

In this paper we focus on the horizon formation when two particles
in 2+1 D collapse head-on. In order to form a true black hole the
particles must have sufficient mass-energy and there must be a
negative cosmological constant. We first recall the equations of
motion for test particles, and we then discuss a model process for
the case of a vanishing cosmological constant. The changes in the
horizon's behavior during the active period for negative
cosmological constant are shown to be relatively minor. We also
mention the collapse of a particle into a black hole and the case
of more than two collapsing particles.

\section{Motion of Test Particles in Anti de~Sitter Space}

A 2+1 dimensional (spinless) point particle is a spacetime with an
angle deficit $\delta$ about the worldline of the particle. It
involves identification by a finite rotation (by the deficit
angle), where the fixed point set (axis) is the particle's
worldline. For particles with general locations this construction
is possible only in homogeneous and isotropic spacetimes, and the
particle's worldline is then necessarily a geodesic. We can
construct such a spacetime from a given background by choosing a
surface (usually totally geodesic timelike) containing the
particle geodesic and rotating it about that geodesic by $\delta$,
remove the wedge volume swept out, and identifying its boundaries
by the rotation.\footnote{The analogous construction using a
spacelike geodesic and Lorentz boost is sometimes referred to as a
tachyon, but it is equally appropriate to interpret it as the
singularity inside a black hole, at least in spacetimes where such
black holes exist.} In the case of several particles (and masses
$\delta$ not so large as to make the spacetime close up), there is
an outer region where the wedges extend outwards from the
particles, and an inner region between the particles where the
background spacetime is unchanged by the cut-and-paste. Within
this inner region the particles therefore move with respect to
each other like geodesic test particles. In flat space ($\Lambda =
0$) this motion is simply the usual constant velocity motion that
occurs when there is no interaction between the particles. In anti
de~Sitter (AdS) space ($\Lambda < 0$), the motion along such
geodesics is not at constant relative velocity but can be easily
derived as follows.

We can choose the geodesic of one particle as an origin. In
``Schwarzschild'' coordinates\footnote{Here the radial
coordinate $q$ is defined so that the circumference of a circle $q =$
const is $2\pi q$. These coordinates are related to another
common form of the AdS metric,
$ds^2 = -\cosh^2r \,dt^2 + dr^2 + \sinh^2r \,d\Omega^2$ by the substitution
 $q = \sinh r$.} the AdS metric (for unit negative curvature, $\Lambda = -1$) is
$$
ds^2 = - (1 + q^2) dt^2 + {dq^2\over 1+q^2} + q^2 d\Omega^2. \eqno{(1)}
$$
(Here $d\Omega^2$ is the metric on the unit sphere; in 2+1 dimensions we simply have $d\Omega^2 = d\phi^2$.) For geodesics with velocity $u^\mu = dx^\mu/d\tau$ the time- and rotational symmetries of (1) imply the conserved quantities
$$
E = u_t = (1 + q^2) {dt\over d\tau} \qquad {\rm and} \qquad L = u_\phi = q^2 {d\phi\over d\tau}
$$
where $E$ and $L$ are the energy and angular momentum per unit mass. In terms of these quantities the normalization condition that $\tau$ be proper time becomes
$$
- {E^2\over 1+q^2} + {1\over 1+q^2}\left({dq\over d\tau}\right)^2
+ {L^2\over q^2} = - 1 \quad {\rm or} \quad \left({dq\over
d\tau}\right)^2 + q^2 + {L^2\over q^2} = E^2 - L^2 - 1.
$$
But this is the same as the ``radial" conservation of energy
equation for a simple harmonic oscillator of unit mass, unit
spring constant, angular momentum $L$, and energy $E^2 - L^2 - 1$,
in terms of the proper time $\tau$. Since the angular equation $L
= q^2 d\phi/d\tau$ is also analogous to that of the 2- (or
higher-) dimensional harmonic oscillator, the motion in $q,\,
\phi$ coordinates, or in rectangular coordinates $x=q\cos\phi, \,
y=q\sin\phi$, is that of a harmonic oscillator: $x$ and $y$ depend
harmonically on $\tau$, the orbits are ellipses, and all geodesics
are periodic with period $2\pi$.

This means in particular that two particles that will in the
future collide head-on at some instant reach a maximum distance
from each other, and their spacetime is time-symmetric about this
instant. Without loss of generality we may therefore assume that
the initial state of such a colliding particle spacetime is
time-symmetric, that is, a two-dimensional spacelike surface of
constant negative curvature and vanishing extrinsic curvature
(totally geodesic hyperbolic 2-space).

\section{Minkowski Space Model}

Although a flat spacetime does not allow horizons, there are null
surfaces that share many of the properties of AdS horizons. In a
locally Minkowskian space the time-symmetric initial state of two
particles, with zero relative velocity of the particles,
corresponds to a static spacetime. We cut out two wedges, each
having its edge at the location of a particle, and choose them to
be symmetrically oriented with respect to the particles (Fig~1a).
The line joining the initial particles on the initial spacelike
surface (or the plane joining their two geodesics in spacetime)
then divides the space into two congruent halves, and the full
space is obtained by ``doubling'' one of the halves, that is
gluing two copies along the edges (Fig~1b). This is rather like a
Melitta coffee filter as it comes out of the box and before it has
been spread into a ``cone.'' The subsequent figures are understood
to be doubled in this way. In the region outside the particles the
initial surface (and all other surfaces $t$ = const) is exactly
conical; we will call it outer-conical, the 2D analog of 3D
asymptotically flat.

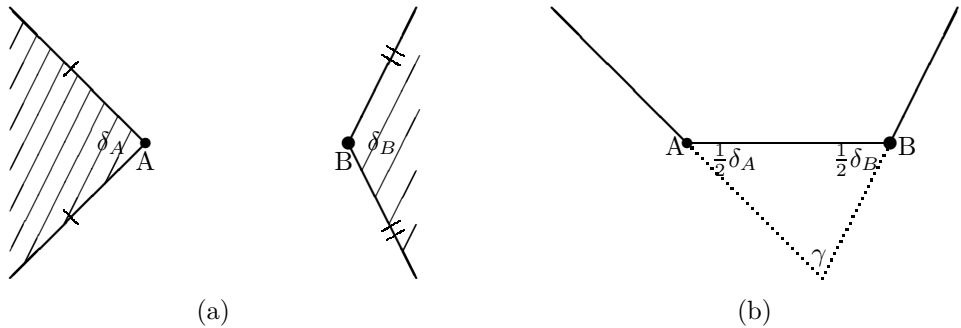
\begin{figure} \label{melitta}

\unitlength 0.90mm
\begin{picture}(150.00,50.00)(10,0)
\put(30.00,30.00){\circle*{1.50}}
\put(60.00,30.00){\circle*{2.00}} \thicklines
\put(60.00,30.00){\line(1,2){10.00}}
\put(30.00,30.00){\line(-1,1){20.00}}
\put(110.00,30.00){\circle*{1.50}}
\put(140.00,30.00){\circle*{2.00}}
\put(140.00,30.00){\line(1,2){10.00}}
\put(110.00,30.00){\line(-1,1){20.00}}
\put(60.00,30.00){\line(1,-2){10.00}}
\put(30.00,30.00){\line(-1,-1){20.00}} \thinlines
\put(110.00,30.00){\line(1,0){30.00}}
\put(30.00,28.67){\makebox(0,0)[ct]{A}}
\put(59.33,28.67){\makebox(0,0)[ct]{B}}
\put(25.00,30.00){\makebox(0,0)[cc]{$\delta_A$}}
\put(65.00,30.00){\makebox(0,0)[cc]{$\delta_B$}}
\put(108.20,31.00){\makebox(0,0)[ct]{A}}
\put(142.50,31.00){\makebox(0,0)[ct]{B}}
\bezier{12}(18.00,40.00)(19.00,41.00)(20.00,42.00)
\bezier{12}(65.00,43.00)(66.25,42.25)(67.50,41.50)
\bezier{12}(65.67,44.33)(66.92,43.58)(68.17,42.83)
\bezier{12}(18.00,20.00)(19.00,19.00)(20.00,18.00)
\bezier{12}(65.00,17.00)(66.25,17.75)(67.50,18.50)
\bezier{12}(65.67,15.67)(66.92,16.42)(68.17,17.17) \thicklines
\bezier{27}(110.00,30.00)(120.00,20.00)(130.00,10.00)
\bezier{22}(130.00,10.00)(135.00,20.00)(140.00,30.00)
\put(117.00,27.50){\makebox(0,0)[cc]{${1\over 2}\delta_A$}}
\put(135.00,27.50){\makebox(0,0)[cc]{${1\over 2}\delta_B$}}
\put(129.50,13.00){\makebox(0,0)[cc]{$\gamma$}}
\put(40.0,5.00){\makebox(0,0)[cc]{(a)}}
\put(120.0,5.00){\makebox(0,0)[cc]{(b)}} \thinlines
\put(12.00,48.00){\line(-1,-2){2.00}}
\put(14.00,46.00){\line(-1,-2){4.00}}
\put(16.00,44.00){\line(-1,-2){6.00}}
\put(18.00,42.00){\line(-1,-2){8.00}}
\put(20.00,40.00){\line(-1,-2){10.00}}
\put(22.00,38.00){\line(-1,-2){12.00}}
\put(24.00,36.00){\line(-1,-2){12.00}}
\put(26.00,34.00){\line(-1,-2){8.00}}
\put(28.00,32.00){\line(-1,-2){4.00}}
\put(62.00,26.00){\line(1,2){8.50}}
\put(64.00,22.00){\line(1,2){6.50}}
\put(66.00,18.00){\line(1,2){4.00}}
\put(68.00,14.00){\line(1,2){2.00}}
\end{picture}

\caption{ (a) A less massive and a more massive static particle, A
and B, on the $t = 0$ spacelike surface of three-dimensional
Minkowski space. The crosshatched wedges are removed from the
space, and the boundaries with equal number of strokes are
identified. The {\it deficit angles} $\delta_A, \, \delta_B$
measure the particle masses. (b) Half of Figure~(a) is sufficient
to represent the space if it is understood that it represents two
layers that are glued together at the boundary -- the result of
folding Figure~(a) along the line AB and identifying the
boundaries as indicated. The total angle deficit is $\delta_T =
2\pi - 2\gamma$.}

\end{figure}

From the figure and the angle sum of the plane triangle at the
bottom of Figure~1b (dotted) it follows immediately that the angle
deficit $\delta_T$ measured at large distances for the equivalent
single particle is the sum of the angle deficits of the two
constituent particles. So the particle masses simply add, as
expected when there is no gravitational interaction energy. The
location of the single particle (tip of the cone) is of course
outside of the 2-particle spacetime. Only in the limit of small
masses is it approximately at the center of mass between them.

Such a spacetime, of course, cannot represent a black hole, since
null geodesics from all events eventually reach infinity. However,
one can consider a congruence of null geodesics, and the
wavefronts normal to them at each Minkowski time, that reach
infinity in the same way that the wavefront of a black hole's
horizon reaches infinity: as a smooth curve of constant curvature
without self-intersection. We will call such a congruence a
pseudo-horizon. Although in flat space such a wavefront can be
arbitrarily translated, this is not the case in a conical space.
The smooth wavefront having any (sufficiently small) constant
curvature in an outer-conical space is {\it unique}. The center of
these wavefronts is the tip of the cone if the space is truly
conical. For outer-conical spaces we can construct the true cone
that analytically continues the outer parts. The wavefronts that
represent the pseudo-horizon at different times are then simply
those parts of circles centered at the tip of this true cone that
lie in the actual space. In the half-figure of the static
two-particle space this tip T is easily constructed by extending
the slanted borders to a point, and the circles representing the
horizon are drawn about this center (Fig~2a). Fig~2b shows both
halves joined at the line between the particles so that successive
stages of the complete pseudo-horizon can be seen. It consists of
constant curvature wavefront sections W that propagate at the
speed of light, interrupted by points of singularities S that
travel at faster-than-light speed toward the particles. New
generators enter the pseudo-horizon along this spacelike line of
singularities. As the pseudo-horizon crosses a particle it becomes
a smooth wavefront (note that the last wavefronts intersect the
boundary of the wedges at right angles). It is clear that the
pseudo-horizon always has the topology of a circle and starts at a
kind of ``center of mass'' P between the particles, or at one of
the particles if that particle has an angle deficit of more than
$\pi$.

\begin{figure}
\unitlength 0.90mm \linethickness{0.4pt}
\begin{picture}(150.00,64.67)(10,-5)
\put(30.00,30.00){\circle*{1.50}}
\put(60.00,30.00){\circle*{2.00}} \thicklines
\put(60.00,30.00){\line(1,2){10.00}}
\put(30.00,30.00){\line(-1,1){20.00}}\thinlines
\put(30.00,30.00){\line(1,0){30.00}}
\put(28.20,31.00){\makebox(0,0)[ct]{A}}
\put(62.50,31.00){\makebox(0,0)[ct]{B}} \thicklines
\bezier{27}(30.00,30.00)(40.00,20.00)(50.00,10.00)
\bezier{22}(50.00,10.00)(55.00,20.00)(60.00,30.00)
\bezier{27}(36.00,24.00)(46.67,33.50)(59.00,28.00)
\bezier{8}(35.00,25.50)(37.67,28.17)(41.67,30.00) \thinlines
\bezier{80}(41.67,30.00)(50.83,34.00)(60.00,30.00)
\bezier{156}(30.00,30.00)(44.50,43.33)(62.50,35.33) \thicklines
\bezier{3}(32.50,27.83)(33.83,29.00)(35.17,30.00) \thinlines
\bezier{120}(35.17,30.00)(48.00,38.50)(61.00,32.17)
\bezier{212}(23.33,36.83)(42.67,54.67)(66.83,43.83)
\put(50.00,9.00){\makebox(0,0)[ct]{T}}
\put(110.00,30.00){\circle*{1.50}}
\put(140.00,30.00){\circle*{2.00}} \thicklines
\put(140.00,30.00){\line(1,2){10.00}}
\put(110.00,30.00){\line(-1,1){20.00}} \thinlines
\put(110.00,30.00){\line(1,0){30.00}}
\put(108.20,31.00){\makebox(0,0)[ct]{A}}
\put(142.50,31.00){\makebox(0,0)[ct]{B}}
\bezier{80}(121.67,30.00)(130.83,34.00)(140.00,30.00)
\bezier{156}(110.00,30.00)(124.50,43.33)(142.50,35.33)
\bezier{120}(115.17,30.00)(128.00,38.50)(141.00,32.17)
\bezier{212}(103.33,36.83)(122.67,54.67)(146.83,43.83) \thicklines
\put(140.00,30.00){\line(1,-2){10.00}}
\put(110.00,30.00){\line(-1,-1){20.00}} \thinlines
\bezier{80}(121.67,30.00)(130.83,26.00)(140.00,30.00)
\bezier{156}(110.00,30.00)(124.50,16.67)(142.50,24.67)
\bezier{120}(115.17,30.00)(128.00,21.50)(141.00,27.83)
\bezier{212}(103.33,23.17)(122.67,5.33)(146.83,16.17)
\put(130.00,35.00){\makebox(0,0)[cc]{W}}
\put(130.00,25.00){\makebox(0,0)[cc]{W}}
\put(130.00,30.00){\makebox(0,0)[cc]{P}}
\put(116.00,30.00){\makebox(0,0)[cc]{S}}
\put(45.00,0.00){\makebox(0,0)[cc]{(a)}}
\put(125.00,0.00){\makebox(0,0)[cc]{(b)}}
\put(92.00,48.00){\line(-1,-2){2.00}}
\put(94.00,46.00){\line(-1,-2){4.00}}
\put(96.00,44.00){\line(-1,-2){6.00}}
\put(98.00,42.00){\line(-1,-2){8.00}}
\put(100.00,40.00){\line(-1,-2){10.00}}
\put(102.00,38.00){\line(-1,-2){12.00}}
\put(104.00,36.00){\line(-1,-2){12.00}}
\put(106.00,34.00){\line(-1,-2){8.00}}
\put(108.00,32.00){\line(-1,-2){4.00}}
\put(142.00,26.00){\line(1,2){8.50}}
\put(144.00,22.00){\line(1,2){6.50}}
\put(146.00,18.00){\line(1,2){4.00}}
\put(148.00,14.00){\line(1,2){2.00}}
\end{picture}

\caption{Construction of pseudo-horizon for two static particles in
flat space. (a) A wavefront that will be smooth and of constant
curvature at late times starts at the tip T of the cone that is
the continuation of the two-particle space's outer region.  (b)
Figure~(a) doubled so that the full shape of the wavefronts can be
seen.}
\end{figure}
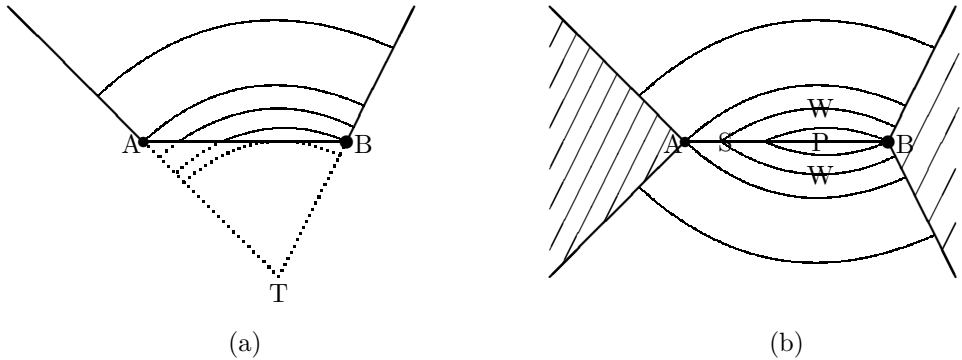

\section{Two particles in Anti de~Sitter space}

Anti de~Sitter  space, like Minkowski space, can be represented in
static coordinates, Eq~(1), as a time sequence of
three-dimensional homogeneous spaces, which however have constant
negative curvature. A point particle centered at the origin can be
constructed as in Minkowski space by removing a timelike wedge of
angle $\delta$ and identifying, for example $\phi = 0$ and $\phi =
2\pi - \delta$. If we define a re-scaled angle $\varphi =
(1-\delta/2\pi)^{-1}\phi$, with the usual $2\pi$ periodicity, and
a re-scaled radial measure $r = (1-\delta/2\pi)q$, the metric (1)
becomes
$$
ds^2 = - (r^2-m) \,dt^2 + {dr^2\over r^2-m} + r^2 \,d\varphi^2 \eqno(2)
$$
with the parameter $m = -(1-\delta/2\pi)^2 < 0$.

These spacelike surfaces are conveniently represented by
Poincar\'e disks, yielding what has been called the ``sausage
model'' of AdS spacetime. In terms of the polar coordinates $\rho,
\phi$ of the disk (related to $q$ by $q = {2\rho\over 1-\rho^2},
\, \rho \leq 1$) the spacetime metric has the ``sausage
coordinate" \cite{holst:28} form
$$
ds^2 = -\left({1+\rho^2\over 1-\rho^2}\right)^2 dt^2 + {4\over (1-\rho^2)^2} (d\rho^2 + \rho^2 d\phi^2) .
$$

Our Minkowski space construction involved mainly straight lines,
planes, and circles. The analogous objects in AdS space are
geodesics, totally geodesic surfaces, and constant curvature
curves. Geodesics and constant non-zero curvature curves are
represented on the Poincar\'e disk by circles perpendicular and
oblique to the boundary of the disk, respectively. Totally geodesic surfaces
intersect the Poincar\'e disks in geodesics, and, in the time
direction, they execute the periodic motion described in section 2.

A two-particle spacetime modeled on AdS space will in general be
dynamic; we saw that this is so even in the test particle limit.
As we also saw, for particles with angular momentum $L = 0$ there
is always a time-symmetric moment, and we will orient our sausage
model so that this moment is one of the spacelike coordinate
surfaces, say $t = 0$. As in Minkowski space we cut the spacetime
into two congruent halves along the geodesics joining the
particles. The initial state of one half will look like Figure~3a,
the AdS analog of Figure~1b, provided the particles' masses are
not too large. (An advantage of the Poincar\'e disk representation
is its conformal nature, so that angle deficits are shown
faithfully. In the alternative ``Klein disk'' representation the
figure at time symmetry would be indistinguishable from a
Minkowski space figure, except for the disk boundary, because
geodesics appear as straight lines on the Klein disk although angles
are distorted.) The equivalent single particle has no location in
the two-particle spacetime, but its angle deficit can be defined
from the behavior of the geometry at large distances. Hyperbolic
trigonometry on the bottom triangle of Figure~3a again shows that
the mass $M$ of the composite particle in terms of the masses
$m_1,\,m_2$ of the constituents and their separation $d$ is given
by
$$
\cos M = \cos m_1 \cos m_2 + \sin m_1 \sin m_2 \cosh d. \eqno(3)
$$
The presence of the factor $\cosh d$ may be viewed as the effect
of the gravitational interaction energy between the particles.

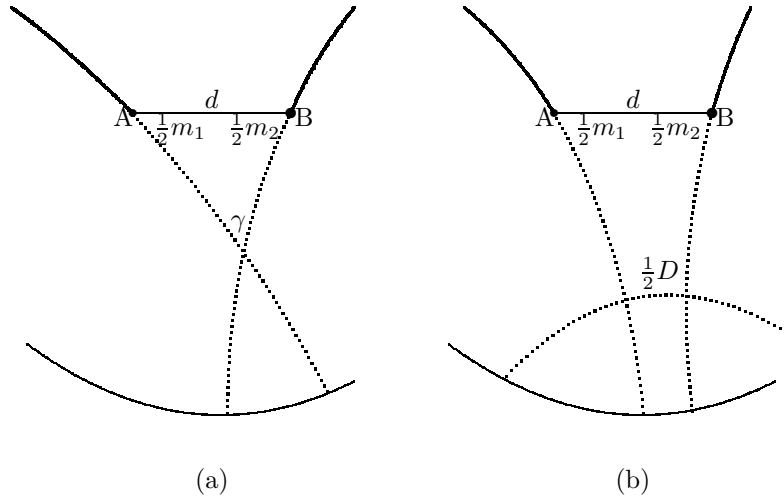
\begin{figure}
\unitlength 0.7mm \linethickness{0.4pt}
\begin{picture}(148.00,95.00)(-27,0)
\put(105.00,75.00){\circle*{1.50}}
\put(135.00,75.00){\circle*{2.00}}
\put(105.00,75.00){\line(1,0){30.00}}
\put(103.20,76.00){\makebox(0,0)[ct]{A}}
\put(137.50,76.00){\makebox(0,0)[ct]{B}} \thicklines
\bezier{80}(62.00,22.00)(38.67,66.33)(5.00,93.00)
\bezier{75}(43.00,18.00)(43.00,63.33)(67.00,95.00)
\bezier{124}(25.00,75.00)(10.67,89.00)(2.00,95.00)
\bezier{96}(55.00,75.00)(59.00,85.00)(67.00,95.00) \thinlines
\bezier{292}(85.00,31.00)(116.00,8.33)(147.00,24.00)
\put(120.00,76.00){\makebox(0,0)[cb]{$d$}}
\put(109.00,72.00){\makebox(0,0)[lc]{${1\over 2}m_1$}}
\put(133.00,72.00){\makebox(0,0)[rc]{${1\over 2}m_2$}}
\put(45.00,54.00){\makebox(0,0)[cc]{$\gamma$}}
\put(25.00,75.00){\circle*{1.50}}
\put(55.00,75.00){\circle*{2.00}}
\put(25.00,75.00){\line(1,0){30.00}}
\put(23.20,76.00){\makebox(0,0)[ct]{A}}
\put(57.50,76.00){\makebox(0,0)[ct]{B}}
\bezier{292}(5.00,31.00)(36.00,8.33)(67.00,24.00)
\put(40.00,76.00){\makebox(0,0)[cb]{$d$}}
\put(29.00,72.00){\makebox(0,0)[lc]{${1\over 2}m_1$}}
\put(53.00,72.00){\makebox(0,0)[rc]{${1\over 2}m_2$}} \thicklines
\bezier{50}(105.00,75.00)(118.67,51.33)(122.00,18.00)
\bezier{108}(88.00,95.00)(97.67,87.00)(105.00,75.00)
\bezier{84}(135.00,75.00)(138.00,85.67)(142.33,95.00)
\bezier{50}(96.00,25.00)(120.00,50.33)(148.00,34.00)
\bezier{50}(131.33,18.33)(127.67,47.33)(135.00,75.00) \thinlines
\put(125.00,41.67){\makebox(0,0)[cb]{${1\over 2}D$}}
\put(40.00,5.00){\makebox(0,0)[cc]{(a)}}
\put(120.00,5.00){\makebox(0,0)[cc]{(b)}}
\end{picture}

\caption{Two particles in anti de~Sitter space at the moment of
their maximum separation $d$, shown on the Poincar\'e disk. The
deficit angles are labeled by the corresponding masses. The disk
is larger than the figure; only its bottom boundary is shown. (a)
In the outer region the geometry is that of a single mass $M =
2\pi - 2\gamma$, which is more than the sum of the individual
masses, corresponding to an increase in interaction energy as the
masses are separated. (b) For larger particle masses or larger
separation the outer geometry is no longer that of a single
particle, but that of a black hole, characterized by its horizon
circumference $D$.}
\end{figure}

The time development in terms of sausage time will show the
approach of the particles to the figure's center and to each
other, as expected from their geodesic motion in the space between
them. They meet there at $t = \pi/2$, and the further time
development cannot be ascertained without a law that determines
the result of this collision (though in the literature some
results are considered more natural than others
\cite{holst:thesis}). Equation~(3) in terms of angles is of course
valid on each Poincar\'e slice of the ``sausage'', but $m_1$ and
$m_2$ are no longer the (constant) rest masses of the particles,
containing a contribution from the kinetic energy. Thus the total
mass $M$ of the system remains unchanged as $d$ changes: at large
distances the system always looks like one static particle, so
there is no horizon and no black hole formation.

It is, however, possible to choose the initial masses and the
distance between them in such a way that Eq~(3) cannot be
fulfilled by a real $M$ because the RHS is greater than 1. In this
case the boundary geodesics that reach infinity (and which
determine $M$ by their angle of intersection, if they do
intersect) are ultraparallel, as in Figure~3b). In terms of an
angle coordinate with the usual $2\pi$ periodicity and a
``Schwarzschild'' radial coordinate that gives the circumference
of circles $r =$ const the usual value $2\pi r$, the metric in the
outer region takes the form (3) with a positive $m$. This is the
BTZ metric \cite{banados:69}, \cite{banados:48} for a non-rotating
black hole, and $m$ is its mass measured asymptotically. (By
contrast, the particle masses $m_1$ and $m_2$ determined by
deficit angles are measured locally.) Analogous to Equation~(3) we
have\footnote{If one or both masses are replaced by black holes, a
similar formula holds with trigonometric functions replaced by
hyperbolic ones.}
$$\cosh D = - \cos m_1 \cos m_2 + \sin m_1 \sin m_2 \cosh d \eqno(4)$$
where $D = 2\pi m$ is the circumference of the black hole throat
at $r = m$. This is twice the minimum distance between the
ultraparallel boundaries in Fig~2b. Thus the problem, so difficult
in 3+1 dimensions, to determine whether given initial data will
lead to black hole formation, is easily solved by considering the
asymptotic dependence of the initial geometry and determining
whether the total BTZ mass $m$ is positive or negative.

If the metric in the outer region is that of a black hole, the
initial position of the horizon is at the throat of the single
black hole described by Eq~(4). Since the outer region covers only
a part of this equivalent black hole, its throat will generally
not be part of the initial space, but as before we can show it by
a kind of analytic continuation as in Figure~3b. On later time
slices, which are superimposed on the initial slice in Figure~4,
the horizon expands and the particles fall toward each other. The
horizon enters the real spacetime at a point on the geodesic
between the particles and spreads out with generally two singular
points on that geodesic, which become smooth when the horizon
crosses the particles. Thus this behavior is qualitatively similar
to that of the Minkowski space in Figure~2a. A major difference is
that the size of the horizon becomes constant once it has passed
all the particles.

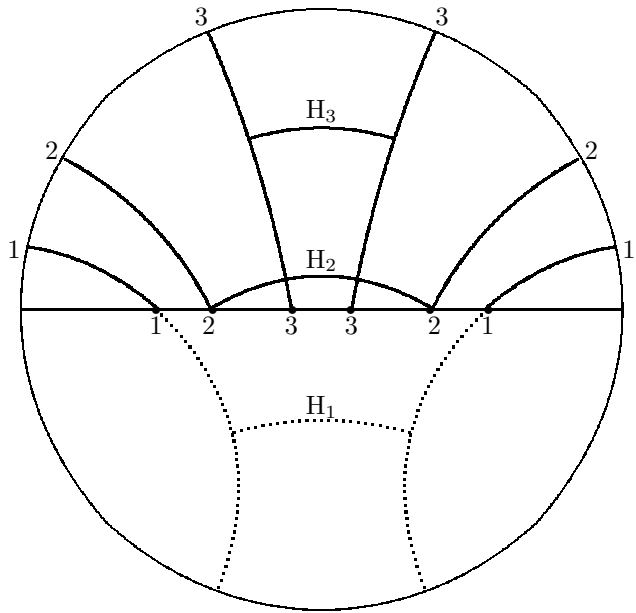
\begin{figure}
\unitlength 1.00mm \linethickness{0.4pt}
\begin{picture}(90.00,90.00)(-20, 5)
\put(10.00,50.00){\line(1,0){80.00}} \thicklines
\bezier{80}(71.67,50.00)(79.67,57.00)(89.00,58.33)
\bezier{116}(64.67,50.00)(71.00,62.67)(84.00,70.00)
\bezier{45}(63.67,13.00)(55.67,34.33)(72.00,50.00)
\bezier{156}(54.00,50.00)(57.33,69.00)(65.00,86.67)
\bezier{80}(28.33,50.00)(20.33,57.00)(11.00,58.33)
\bezier{45}(36.33,13.00)(44.33,34.33)(28.00,50.00)
\bezier{116}(35.33,50.00)(29.00,62.67)(16.00,70.00)
\bezier{156}(46.00,50.00)(42.67,69.00)(35.00,86.67)
\bezier{24}(38.33,33.61)(50.00,36.94)(61.67,33.61)
\bezier{140}(35.00,50.00)(50.00,58.9)(65.00,50.00)
\bezier{80}(40.56,72.78)(50.00,75.56)(59.44,72.78) \thinlines
\bezier{132}(10.00,50.00)(10.00,65.28)(21.39,78.33)
\bezier{132}(21.39,78.33)(33.89,89.72)(50.00,90.00)
\bezier{132}(90.00,50.00)(90.00,65.28)(78.61,78.33)
\bezier{132}(78.61,78.33)(66.11,89.72)(50.00,90.00)
\bezier{132}(10.00,50.00)(10.00,34.72)(21.39,21.67)
\bezier{132}(21.39,21.67)(33.89,10.28)(50.00,10.00)
\bezier{132}(90.00,50.00)(90.00,34.72)(78.61,21.67)
\bezier{132}(78.61,21.67)(66.11,10.28)(50.00,10.00)
\put(28.0,50.00){\circle*{1.11}} \put(35.56,50.00){\circle*{1.11}}
\put(46.11,50.00){\circle*{1.11}}
\put(72.22,50.00){\circle*{1.11}}
\put(64.44,50.00){\circle*{1.11}}
\put(53.89,50.00){\circle*{1.11}}
\put(10.00,58.00){\makebox(0,0)[rc]{1}}
\put(28.00,49.00){\makebox(0,0)[ct]{1}}
\put(35.00,49.00){\makebox(0,0)[ct]{2}}
\put(46.00,49.00){\makebox(0,0)[ct]{3}}
\put(15.00,70.00){\makebox(0,0)[rb]{2}}
\put(34.00,89.00){\makebox(0,0)[cc]{3}}
\put(90.00,58.00){\makebox(0,0)[lc]{1}}
\put(73.00,49.00){\makebox(0,0)[rt]{1}}
\put(65.00,49.00){\makebox(0,0)[ct]{2}}
\put(54.00,49.00){\makebox(0,0)[ct]{3}}
\put(85.00,70.00){\makebox(0,0)[lb]{2}}
\put(66.00,89.00){\makebox(0,0)[cc]{3}}
\put(50.00,37.00){\makebox(0,0)[cc]{H$_1$}}
\put(50.00,56.50){\makebox(0,0)[cc]{H$_2$}}
\put(50.00,76.40){\makebox(0,0)[cc]{H$_3$}}
\end{picture}
\caption{Three time slices in sausage coordinates of two-particle
collapse and associated horizon, superimposed. The outer circle
represents infinity. Only the half space is shown, the complete
configuration at each time is obtained by reflection about the
horizontal line and identifying the heavy curves that go to
infinity. The two equal-mass particles are indicated by black dots
at successive times 1, 2, 3. The initial, time-symmetric
configuration is represented by the curves between by the four
points labeled 1. There is no horizon initially, but the outer
geometry, when continued inward without particle singularities, is
shown by the dotted curves and would have a horizon at H$_1$. At
the time labeled 2, the particles have approached each other, and
the horizon H$_2$ has already propagated into the actual space and
just reached the particles. In the third configuration, the
particles are closer to collision, and the horizon $H_3$ surrounds
the particles, propagating outward toward infinity. The horizon's
circumference remains constant after the position H$_2$, though
the Poincar\'e disk representation does not show its true size.}
\end{figure}

\section{Generalizations}

The starting ``point'' of the horizon is not just the point on the
line between the particles where it first appears in sausage time.
Every point on that line contributes at some time a pair of
generators (one for each half space) to the horizon. Because the
horizon moves faster than the speed of light along this line, the
line of the horizon's origin in spacetime is spacelike. In a
suitable, different time slicing the horizon may therefore appear
simultaneously everywhere along the line, and then spread out in
all directions away from it \cite{mats}, or it may start at one of
the masses and have a singularity at only one point, which then
runs to the other mass. (In fact, this is the case in sausage time
if one particle's mass is sufficiently larger than the other's.)

\begin{figure}
\unitlength 0.80mm \linethickness{0.4pt}
\begin{picture}(141.00,83.57)(-11,0)
\thicklines \put(59.47,40.00){\line(1,0){19.0}}
\put(38.67,51.33){\circle*{1.33}}
\bezier{116}(38.67,51.33)(28.00,61.67)(16.33,69.00)
\bezier{92}(38.67,51.33)(27.00,53.00)(15.67,55.67)
\bezier{96}(38.67,51.33)(50.33,46.7)(59.33,40.00)
\put(38.67,28.67){\circle*{1.33}}
\bezier{116}(38.67,28.67)(28.00,18.33)(16.33,11.00)
\bezier{92}(38.67,28.67)(27.00,27.00)(15.67,24.33)
\bezier{96}(38.67,28.67)(50.33,33.33)(59.33,40.00)
\put(97.19,40.00){\line(-1,0){18.53}}
\put(118.00,51.33){\circle*{1.33}}
\bezier{116}(118.00,51.33)(128.67,61.67)(140.33,69.00)
\bezier{92}(118.00,51.33)(129.67,53.00)(141.00,55.67)
\bezier{96}(118.00,51.33)(106.33,46.7)(97.33,40.00)
\put(118.00,28.67){\circle*{1.33}}
\bezier{116}(118.00,28.67)(128.67,18.33)(140.33,11.00)
\bezier{92}(118.00,28.67)(129.67,27.00)(141.00,24.33)
\bezier{96}(118.00,28.67)(106.33,33.33)(97.33,40.00) \thinlines
\bezier{108}(65.52,39.90)(78.19,43.90)(90.86,39.90)
\bezier{260}(50.19,45.57)(78.19,62.24)(106.19,45.57)
\bezier{408}(38.52,51.57)(77.86,83.57)(117.52,51.57)
\bezier{44}(50.28,45.56)(49.17,40.00)(50.28,34.44)
\bezier{96}(38.67,51.33)(35.56,40.00)(38.67,28.67)
\bezier{108}(65.52,40.10)(78.19,36.10)(90.86,40.10)
\bezier{260}(50.19,34.43)(78.19,17.76)(106.19,34.43)
\bezier{408}(38.52,28.43)(77.86,-3.57)(117.52,28.43)
\bezier{44}(106.39,45.56)(107.50,40.00)(106.39,34.44)
\bezier{96}(118.00,51.33)(121.11,40.00)(118.00,28.67)
\put(29.67,53.00){\line(1,0){7.17}}
\put(34.67,55.00){\line(-1,0){16.17}}
\put(15.83,57.00){\line(1,0){16.67}}
\put(30.00,59.00){\line(-1,0){14.00}}
\put(16.00,61.00){\line(1,0){11.50}}
\put(25.33,63.00){\line(-1,0){9.33}}
\put(22.33,65.00){\line(-1,0){6.33}}
\put(16.00,67.00){\line(1,0){3.33}}
\put(29.67,27.00){\line(1,0){7.17}}
\put(34.67,25.00){\line(-1,0){16.17}}
\put(15.83,23.00){\line(1,0){16.67}}
\put(30.00,21.00){\line(-1,0){14.00}}
\put(16.00,19.00){\line(1,0){11.50}}
\put(25.33,17.00){\line(-1,0){9.33}}
\put(22.33,15.00){\line(-1,0){6.33}}
\put(16.00,13.00){\line(1,0){3.33}}
\put(127.67,53.00){\line(-1,0){7.17}}
\put(122.67,55.00){\line(1,0){16.17}}
\put(141.50,57.00){\line(-1,0){16.67}}
\put(127.33,59.00){\line(1,0){14.00}}
\put(141.33,61.00){\line(-1,0){11.50}}
\put(132.00,63.00){\line(1,0){9.33}}
\put(135.00,65.00){\line(1,0){6.33}}
\put(141.33,67.00){\line(-1,0){3.33}}
\put(127.67,27.00){\line(-1,0){7.17}}
\put(122.67,25.00){\line(1,0){16.17}}
\put(141.50,23.00){\line(-1,0){16.67}}
\put(127.33,21.00){\line(1,0){14.00}}
\put(141.33,19.00){\line(-1,0){11.50}}
\put(132.00,17.00){\line(1,0){9.33}}
\put(135.00,15.00){\line(1,0){6.33}}
\put(141.33,13.00){\line(-1,0){3.33}}
\end{picture}

\caption{Development of the horizon in the collapse of four
particles. The picture for each of the three times at which the
horizon is shown was enlarged so that the particles appear at
constant positions, but it is schematic only in certain respects.
The heavy, V-shaped lines in the outer parts are to be identified
as in previous figures to create the angle deficits. The inner
heavy lines show the paths of the singular points. The lighter
curves are stages of the horizon up to the time when it reaches
the particles.}
\end{figure}
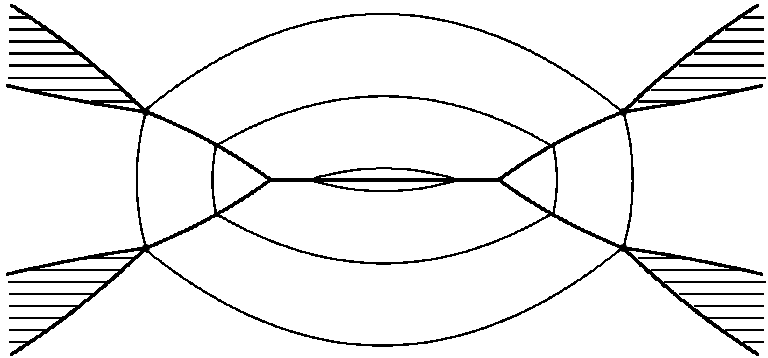

In our figures we can replace the geodesics that intersect at one
or both of the particles by ultraparallel ones. Such initial
states can be described as a particle falling into a black hole,
or two black holes merging. For two black holes the horizon of the
final single black hole is always present on an initially
time-symmetric surface [6, 7], and for a black hole and a particle
it is at least partially exposed. In the latter case the initial
horizon is not that of the black hole alone, but it has a
singularity and is thus prepared to become smooth by swallowing
the particle. For two black holes (or more extreme cases of black
hole and particle) the horizon's future time development is simply
that of the resulting single black hole, with no trace of it
having arisen from a collapse. However, in the past of the
time-symmetric initial surface the horizon consisted of two
separate circular parts, each of which had singularities that
moved toward each other and cancelled after they merged into a
single circle.

If there are more than two particles collapsing to form a black
hole, the main qualitative difference is the pattern formed by the
horizon singularities. Let us follow the smooth, constant
curvature horizon of some late time backwards in time. It
contracts and remains smooth at increasing curvature until it
crosses a particle. After crossing it acquires a discontinuity
equal to the particle's angle deficit. Such discontinuities
propagate and increase inward along geodesics from each particle.
The lines of discontinuity meet in pairs and merge until a single
line with two singularities propagating toward each other that
eventually annihilate. Thus the branching tree pattern of the
singularities tell the essential story of the horizon's
development. An example for the case of four particles is shown in
Figure~5.

If the point particles are replaced by finite but concentrated
matter distributions, the singular points of the horizon on
spacelike surfaces will be replaced by concentrations of high
extrinsic curvature, running along a similar tree pattern.

In the case that the particles have non-zero angular momentum, they
never meet and instead follow periodic orbits in the space between
them. With appropriate initial conditions they can, nevertheless,
form a black hole, if one follows the usual custom in 2+1 D AdS
geometries to regard as singular only the region of certain smooth
timelike curves. This region has a gap through which the particles
can pass at closest approach and then separate again, albeit into
another universe, as is the case for the analytic extension of the
Kerr geometry. A case of black holes on circular orbits has been
discussed by  DeDeo and Gott \cite{dedeo}, and similar configurations have
been recognized as a rotating BTZ wormhole by Holst and Matschull
\cite{holst:16}. It will be interesting to explore the tree pattern of the
horizon singularities in such cases.

\end{document}